\documentclass[12pt]{article}
\usepackage{epsfig}
\usepackage{amsfonts}
\usepackage{amssymb}
\usepackage{indentfirst}
\usepackage{color}
\usepackage{mathrsfs}
\def\bc{\begin{center}}

\def\ec{\end{center}}
\def\be{\begin{eqnarray}}
\def\ee{\end{eqnarray}}

\definecolor{dyellow}{rgb}{1.,0.8,.0}
\definecolor{myblue}{rgb}{.1,.1,.7}
\definecolor{dcyan}{rgb}{.0,.6,.6}
\definecolor{dmagenta}{rgb}{0.6,0.0,0.6}
\definecolor{brown}{rgb}{0.6,0.2,0.}
\definecolor{darkblue}{rgb}{.0,.0,0.5}
\definecolor{darkred}{rgb}{0.75,0.0,0.0}
\definecolor{orange}{rgb}{1.,.6,.0}
\definecolor{dorange}{rgb}{0.8,.4,.0}
\definecolor{darkgreen}{rgb}{0.0,0.6,0.0}
\definecolor{purple}{rgb}{.4,.0,.4}

\begin{document}
\baselineskip=16pt
\newcommand{\omits}[1]{}

\hfill{\bf BIHEP-TH-2005-7} \vspace{1.0cm}

\begin{center}
{\Large \bf Cosmic ray threshold \\ in an asymptotically dS
spacetime
}\\
\vspace{1cm}
Zhe Chang\footnote{changz@mail.ihep.ac.cn}~~and~~Shao-Xia Chen\footnote{ruxanna@mail.ihep.ac.cn}\\
{\em Institute of High Energy Physics,
Chinese Academy of Sciences} \\
{\em P.O.Box 918(4), 100049 Beijing, China}\\
Cheng-Bo Guan\footnote{guancb@ustc.edu.cn}\\
{\em Interdisciplinary Center for Theoretical Study}\\
{\em University of Science and Technology of China, 230026 Hefei, China}\\
Chao-Guang Huang\footnote{huangcg@mail.ihep.ac.cn} \\
{\em Institute of High Energy Physics,
Chinese Academy of Sciences} \\
{\em P.O.Box 918(4), 100049 Beijing, China}\\
\end{center}
\vspace{1.5cm}
\begin{abstract}
We discuss the threshold anomaly in ultra-high energy cosmic ray
physics by assuming that the matter world just be perturbation of
a de Sitter universe, which is consistent with the recent
astronomical observations: about two thirds of the whole energy in
the universe is contributed by a small positive cosmological
constant. One-particle states are presented explicitly. It is
noticed that the dispersion relation of free particles is
dependent on the degrees of freedom of angular momentum. This fact
can be regarded as the effects of the cosmological constant on
kinematics of particles.
\end{abstract}

 \vspace{1.0cm}

\section{Introduction}
The origin of the ultra high energy cosmic rays (UHECRs) is one of
the outstanding puzzles of modern astrophysics. Present
understanding of the phenomena responsible for the production of
UHECR is still limited. Currently, there are generally two
scenarios to produce the UHECRs. One is the ``bottom-up"
acceleration scenario with some astrophysical objects as
sources\cite{acc1,acc2}. The other is called ``top-down" scenario
in which UHECR particles are from the decay of certain
sufficiently massive particles originating in the early
Universe\cite{top1}.

Decades ago, Greisen, Zatsepin and Kuz'min (GZK) \cite{GZK}
discussed the
propagation of 
UHECRs through the cosmic microwave background radiation (CMBR).
Due to photopion production process by the CMBR, the UHECR
particles will lose their energies drastically down to a
theoretical threshold, which is about
$5\times 10^{19}$eV. 
The mean free path for this process is only a few Mpc \cite{GZK1}.
This is the so-called GZK cutoff. However, we have observed indeed
hundreds of events with energies above $10^{19}$eV and about 20
events above $10^{20}$eV \cite{data1}-\cite{data5}. The puzzle can
be considered to be some cosmic ray threshold anomaly: energy of
an expected threshold is reached but the threshold has not been
observed yet. Although some physicists  argued that the GZK cutoff
might actually have already been seen \cite{0206217}, it need more
evidences and careful analysis to become the consensus in
scientific circles. In this paper, we will take the existence of
UHECR paradox as an acceptable working assumption, on the basis of
which we try to unravel the puzzle.

The cosmic-ray paradox was suggested to be solved by the
departures from Lorentz symmetry in the end of the last century
\cite{glashow1, glashow2}.  The basic idea is to introduce tiny
Lorentz-violating terms into the standard model Lagrangian and
derive a parameter-dependent threshold.  As long as the parameters
are taken some appropriate values, the threshold will be above the
energies of all observed events.  A related but alternative scheme
is proposed \cite{planck}, in which Lorentz symmetry is broken by
Planck-scale effects. Another approach is to assume deformed
Lorentz symmetry, constructing the theory of doubly special
relativity (DSR) \cite{DSR1, DSR2, DSR3}.  In the approach, a new
dispersion relation is written down and then an enhanced threshold
is obtained.

The recent astronomical observations on supernovae
\cite{constant-super-1, constant-super-2} and CMBR
\cite{costant-cmbr} show that about two thirds of the whole energy
in the Universe is contributed by dark energy.  The simplest model
for dark energy is that it is devoted by a small positive
cosmological constant $(\Lambda)$. Then, the Universe can be
regarded as a de Sitter (dS) spacetime in the zeroth
approximation. The physics in dS spacetime has been discussed
extensively (see, for example,
\cite{ds-paper-1}-\cite{ds-paper-3}).

In our previous paper \cite{0307439}, we have discussed the
possibility of the cosmological constant as the origin of
threshold anomaly in a dS spacetime. In a kind of simplified case,
we obtained a positive conclusion. However there are two
imperfects in Ref. \cite{0307439}, which are also motivations for
this paper. Firstly, although the dominated ingredient in the
Universe is dark energy, the matter also contributes about 1/3 to
the density of the Universe. The challenging task is to find a way
of dealing with the universe dominated by the dark energy and
matter. At least at the present stage of the evolution of the
universe, one has few reasons to omit effects of any of the two
ingredients. Therefore, we need to investigate the Universe
including matter in order to more approach to the Universe
nowadays. Secondly, in \cite{0307439}, only some simplified cases
are considered, which are not necessary the true ones in our
Universe and will be improved on in current paper.

In this paper, we present a scenario in which the dark energy is
supposed to come from the cosmological constant and that the
spacetime is a dS one, and the matter in the Universe is dealt
with as a perturbation around the dS background. The perturbation
of the spacetime geometry is derived firstly. Then we discuss the
motion of a free particle in the asymptotically de Sitter
Universe, and its kinetics is set up primarily. Meanwhile, we
derive a general form of dispersion relation for free particles
moving in the Universe. This formalism is used to investigate the
UHECR propagating in the cosmic microwave background. We obtain
explicitly the corrections of the GZK threshold for the UHECR
particles interacting with soft photons, which are dependent on
the cosmological constant as supposed in the beginning of the
paper. We show how the threshold varies with a positive
cosmological constant and additional degrees of freedom of the
angular momentums of interacting particles. It should be noticed
that, for a positive cosmological constant, the theoretic
threshold tends to be above the energies of all the observed
events. Thus, we may conclude that the tiny but nonzero
cosmological constant is a possible origin of the threshold
anomaly of the UHECR.

The paper is organized as follows. In Section 2, we discuss the
Friedmann equation for a constant curvature spacetime with a
homogeneous density perturbation. An explicit solution including
metric and corresponding connection is presented. The section 3 is
devoted to the investigation of kinematics. Approximate
conservation laws of momentum and angular momentum are obtained
along the geodesics. By solving equations of motion of a free
particle, we obtain a remarkable dispersion relation, which
includes degrees of freedom of angular momentum. In Section 4, by
taking effects of a tiny but nonzero positive cosmological
constant into account, we show that the theoretic threshold is
above the energies of all the observed UHECR events. In the last
section, we present conclusions and remarks.

\section{Perturbation of the Friedmann equations}
The Einstein equation with cosmological constant is
\begin{equation}
\label{einstein} R_{ab}-\frac{1}{2}g_{ab}R-\Lambda g_{ab}=-8\pi
GT_{ab}~.
\end{equation}
The stress-energy tensor in Eq. (\ref{einstein}) can be of the
general perfect fluid form
\begin{equation}
T_{ab}=(\rho+p)U_{a} U_{b}-pg_{ab}~,
\end{equation}
where $p,~\rho$ are the proper pressure and energy density,
respectively.

While observations of the distribution of galaxies in our Universe
show clustering of galaxies on a wide range of distance scales, on
the largest scales the galaxy distribution appears to be
homogeneous and isotropic, namely, $p$ and $\rho$ are only
functions of variable $t$. The homogeneous and isotropic universe
is described by the Robertson-Walker metric
\begin{equation}
\label{FRW-ds}
ds^2=dt^2-\hat{a}^2(t)\left[\frac{dr^2}{1-kr^2}+r^2(d\theta^2+\sin^2\theta
d\phi^2)\right]~,
\end{equation}
where $k$ denotes spatial curvature of the spacetime.

Thus, the general evolution equations for homogenous, isotropic
universe read
\begin{equation}
\begin{array}{c}
 3 \ddot{\hat{a}}=-4\pi G(\rho+3p)\hat{a}~\\[0.21cm]
 \hat{a}\ddot{\hat{a}}+2\dot{\hat{a}}^2+2k=4\pi G(\rho-p)\hat{a}^2~,
 \end{array}
 \end{equation}
where $\rho=\rho_{\Lambda}+\rho_m$, $\rho_{\Lambda}$ and $\rho_m$
denote the energy density contributed from cosmological constant
and matter, respectively.

The stress-energy tensor should be covariantly conserved, {\it
i.e.}, $T^{ab}_{~~;b}=0$, which reduces to
 \begin{equation}
 \dot{\rho}=-3\frac{\dot{\hat{a}}}{\hat{a}}(\rho+p)~.
 \end{equation}
It should be noticed that $\rho_m/\rho_{\Lambda}$ is a small
parameter for the evolution of the Universe (at least in the
present stage). To discuss deviation of physical quantities in an
asymptotically de Sitter spacetime from those in a dS one, we
would introduce a small parameter $\epsilon$ characterizing the
effect of the matter on the spacetime geometry. Then, one can
assume the scale factor $a(t)$ has a perturbation by the matter
density $\rho_m$ around a dS background,
\begin{equation}
\label{hat-a} \hat{a}(t)=a(t)[1+\epsilon
f(t)]=R\cosh(t/R)[1+\epsilon f(t)]~,
\end{equation}
where $R:=\sqrt{3/\Lambda}$.

From the first approximation of the covariant conservation
equation
\begin{equation}
\label{mass-c} \dot{\rho}+3H(\rho+p)=0~,
\end{equation}
and
\begin{equation}
\label{lambda-m}
\rho=\rho_{\Lambda}+\rho_m=\displaystyle\frac{3}{8\pi
G}\left(\displaystyle\frac{1}{R^2}+\epsilon\rho(t)\right)~.
\end{equation}
one can obtain
\begin{equation}\label{f}
f(t)=\displaystyle\frac{\rho_0R^2}{2}\left( \tanh(t/R)\cot^{-1}
[\sinh(t/R)]-\frac{1}{\cosh(t/R)}\right)~.
\end{equation}
In the derivation, the asymptotic condition $\lim\limits_{t\to
\infty}f(t)\to 0$ has been used.  The asymptotically dS spacetime
with the above form of matter can be realized as a four
dimensional hypersurface embedded in a five dimensional flat space
\be ds^2=(d\hat{\xi}^0)^2-(d\hat{\xi}^1)^2-
(d\hat{\xi}^2)^2-(d\hat{\xi}^3)^2-(d\hat{\xi}^4)^2~, \ee such that
\be \label{hypersurface}
(\hat{\xi}^0)^2-(\hat{\xi}^1)^2-(\hat{\xi}^2)^2-(\hat{\xi}^3)^2
-(\hat{\xi}^4)^2=-\hat{R}^2(t)~. \ee $\hat{\xi}^\mu$
($\mu=0,~1,~2,~3,~4~$) is related to the coordinate
($t,~r,~\theta,~\phi~$) in Eq. (\ref{FRW-ds}) as follows \cite{xu}
\be \label{xi-hat} \hat{\xi}^0   -\hat \xi^0(t=0) =
\displaystyle\int^t_0\sqrt{1+\dot{\hat{a}}^2(\tau)}d\tau ~, \qquad
\hat{\xi}^A=\xi^A(1+\epsilon f)~, \qquad (A=1,2,3,4)~, \ee where
\begin{equation}
\begin{array}{l}
\xi^1=r\cosh(t/R)\sin\theta\cos\phi , \\ [0.3cm]
\xi^2=r\cosh(t/R)\sin\theta\sin\phi ,\\ [0.3cm]
\xi^3=r\cosh(t/R)\cos\theta ,\\ [0.3cm] \xi^4=\sqrt{R^2-r^2}\cosh(t/R) .\\
[0.3cm]
\end{array}
\end{equation}
Applying Eqs. (\ref{hat-a}) and (\ref{f}) to the expression of
$\hat{\xi}^0$ in (\ref{xi-hat}), one obtains \be \label{xi0-hat}
\hat \xi^0 
=\omits{& \fbox{$\xi^0 + \epsilon
\displaystyle\frac{\rho_0R^3}{2}\left(\tanh(t/R)
+\cosh(t/R)\mbox{arc\,ctan}(\sinh(t/R))-2\right)$}  \\ [0.4cm]
&=&\fbox{$\xi^0 + \epsilon R\displaystyle\frac
{\cosh^2(t/R)}{\sinh(t/R)}f(t)
+\epsilon\displaystyle\frac{\rho_0R^3}{2} \left( {\tanh(t/R)}
+\mbox{coth}(t/R)-2\right) $}\\ [0.4cm] &=&}
\xi^0 + \epsilon
R\displaystyle\frac{\cosh^2(t/R)}{\sinh(t/R)}f(t) +\epsilon
\rho_0R^3 \left( \mbox{coth}(2t/R)-1\right)~,
\ee where \be \xi^0 = R\sinh(t/R). \ee Similarly, in the course of
deriving Eq. (\ref{xi0-hat}), the condition
$\lim\limits_{t\to\infty}\hat{\xi}^0=\xi^0$ is used. $\hat{R}(t)$
in Eq. (\ref{hypersurface}) reads
\begin{equation} \label{R-1}
\hat{R}(t)=R\left(1-\epsilon\rho_0
R^2\sinh(t/R)(\mbox{coth}(2t/R)-1)\right)~.\end{equation}It is
obvious that
\begin{equation}
\label{R-infty} \lim\limits_{t\to \infty}\hat{R}(t)=R~.
\end{equation}
For convenience, we write the metric in terms of the Beltrami
coordinates $x^{a}= R\displaystyle \frac{\xi^{a}}{\xi^4}~,
(\xi^4\not=0~,a =0,\  1,~2,~3$)
\cite{wulixuebao-1}-\cite{wulixuebao-2}. Due to the additional
matter, the metric of the 4-dimensional spacetime in Beltrami
coordinates has a perturbed form
\begin{equation}
\label{metric}
ds^2=
\hat{g}_{ab}dx^{a}dx^{b}=(g_{ab}+\epsilon
h_{ab})dx^{a}dx^{b}~,
\end{equation}
where $g_{ab}$ is the metric in the empty dS spacetime with the
form
\begin{equation}\label{sigma}
\begin{array}{l}
\displaystyle g_{ab}= \frac{\eta_{ab}}
{\sigma}+\frac{\eta_{ac}\eta_{bd}x^{c}x^{d}}
{\sigma^2R^2}~,~~\eta_{ab}={\rm diag}(1,~-1,~-1,~-1)~,\\[3mm]
\sigma:=\sigma(x,~x)=1-\displaystyle\frac{\eta_{ab}x^{a}x^{b}}{R^2}>0
\end{array}
\end{equation}
and $h_{ab}$ is the perturbation from the additional matter.

From Eq. (\ref{metric}), the following results can be obtained
\begin{equation}
\label{h-x-a}
\begin{array}{rcl}
h_{0a}&=&0~,\\ [0.5cm] h_{ij} &=&
\omits{\fbox{$\displaystyle\frac{\rho_0R^2}{\sigma}\left(\displaystyle\frac{x^0}{\sqrt{\sigma
R^2+(x^0)^2}}{\rm
arccot}\left(\displaystyle\frac{x^0}{\sqrt{\sigma}R}\right)-\displaystyle\frac{\sqrt{\sigma}R}{\sqrt{\sigma
R^2+(x^0)^2}}\right)\left(\displaystyle\frac{\eta_{ki}x^k\eta_{lj}x^l}{\sigma
R^2+(x^0)^2}-\delta_{ij}\right)~.$} \\ [0.7cm] &=& \fbox{$
\displaystyle\frac{\rho_0R^2}{\sigma}
\frac{\sqrt{\sigma}R}{\sqrt{\sigma R^2+(x^0)^2}}
\left(\displaystyle\frac{x^0}{\sqrt{\sigma} R}{\rm
arccot}\left(\displaystyle\frac{x^0}{\sqrt{\sigma}R}\right)-1\right)\left(\displaystyle\frac{\eta_{ki}x^k\eta_{lj}x^l}{\sigma
R^2+(x^0)^2}-\delta_{ij}\right)~  $}\\ [0.7cm] &=&} - 2 f(t)
\displaystyle\frac{\sigma R^2 +(x^0)^2
}{\sigma(x)R^2}~^3g^{B}_{ij} =- 2
f(t)R^{-2}a^2(t)\;^3g^{B}_{ij}~,~~i,~j=1,~2,~3~;
\end{array}
\end{equation}
where $\,^3g^{B}_{ij}$ is the Beltrami metric on 3-sphere
\begin{equation}
\begin{array}{rcl}
~^3g^{B}_{ij}&=&\displaystyle\frac {\delta_{ij}}{\sigma_3(x)}-\displaystyle\frac{\delta_{ik}\delta_{jl} x^k x^l}
 {\sigma_3^2(x)R^2} , \\ [0.5cm]
\sigma_3(x)&=&1+R^{-2}\delta_{ij}x^ix^j~.
 \end{array}
 \end{equation}

In order to investigate the motion of free particles, Christoffel
coefficients are listed as follows
\begin{equation}\label{connection}
\begin{array}{rcl}
\hat{\Gamma}^{a}_{bc} &=&\Gamma^{a}_{bc} +\epsilon\widetilde{\Gamma}
^{a}_{bc}~, ~~\Gamma^{a}_{bc}=(\eta_{bd}\delta^{a}_{c} +
\eta_{cd}\delta^{a}_{b})\displaystyle\frac{x^d}{R^2 \sigma(x)}; \\
[0.5cm] \widetilde{\Gamma} ^0_{0i} &=& \displaystyle\frac{\rho_0}
{2} \tanh (t/R)\cot^{-1}[\sinh(t/R)] \,
\displaystyle\frac {x^i}{\sigma_3}, \\
[0.5cm]\widetilde{\Gamma}^0_{ij} &=& \displaystyle\frac {\rho_0
R}{2} \{\cosh(t/R)[1+\tanh^2(t/R)]\cot^{-1}[\sinh(t/R)]-2
\tanh(t/R)\} \sigma^{1/2} \; ^3g^B_{ij}  \\ [0.5cm]
&&-\displaystyle\frac {\rho_0 }{R} \tanh^2(t/R)
\cot^{-1}[\sinh(t/R)]\displaystyle\frac {x^ix^j}{\sigma_3^{3/2}} ,
\\  [0.5cm]
\widetilde{\Gamma} ^i_{0j} &=& \displaystyle\frac {\rho_0 R}
{2\sqrt{\sigma_3}} \cot^{-1}[\sinh(t/R)]\, \delta^i_j , \\
[0.5cm] \widetilde{\Gamma}^k_{ij}&=&-\displaystyle\frac
{\rho_0}{2}\tanh(t/R)\cot^{-1}[\sinh(t/R)] \displaystyle\frac{x^i
\delta^k_{j} + x^j \delta^k_{i}}{\sigma_3}.
\end{array}
\end{equation}
It is not difficult to calculate the deviation of the geodesics in
an asymptotically dS spacetime from those in the empty dS one
\begin{equation}\label{geodesics}
0=\displaystyle\frac{d^2x^a}{ds^2}+\hat{\Gamma}^a_{bc}\displaystyle
\frac{dx^b}{ds}\displaystyle\frac{dx^c}{ds}
=
\omits{\displaystyle\frac{d^2x^a}{ds^2}+{\Gamma}^a_{bc}\displaystyle
\frac{dx^b}{ds}\displaystyle\frac{dx^c}{ds}+\epsilon
\tilde{\Gamma}^a_{bc}\displaystyle
\frac{dx^b}{ds}\displaystyle\frac{dx^c}{ds} \\ [0.5cm]
&=&}
\sigma\displaystyle\frac{d}{ds}\left(\displaystyle\frac{1}{\sigma}\displaystyle\frac{dx^a}{ds}
\right)+\epsilon\widetilde{\Gamma}^a_{bc}\displaystyle\frac{dx^b}{ds}
\displaystyle\frac{dx^c}{ds}~.
\end{equation}
In the asymptotically dS spacetime with homogeneous matter, one
can define the 4-momentum formally as
$$\hat{P}^a:=\displaystyle\frac{1}{\sigma}\displaystyle\frac{dx^a}{ds}~,~~~{\rm and}~~\hat{E}:=\hat{P}^0~.$$
From Eq. (\ref{geodesics}), one has
\begin{equation}
\begin{array}{rcl}
\label{energy-perturbation}
d\hat{P}^a=-\epsilon\widetilde{\Gamma}^a_{bc}\hat{P}^bdx^c\Rightarrow
\displaystyle\frac{d\hat{P}^a}{dx^0}=-\epsilon\widetilde{\Gamma}^a_{bc}\hat{P}^bv^c~,
\end{array}
\end{equation}
where $v^c$ is a constant defined as
$$v^a:=\frac{dx^a}{dx^0}~.$$

For the particle moving along a one-dimensional curve, one can set
$x^2=x^3=0$ without loss of generality. That is, one can discuss
the motion of a free particle in $x^0-x^1$ plane. In this case,
the deviations of energy and momentum in the perturbed dS universe
from those in the empty dS universe can be calculated numerically.
In Fig. 1, we present a plot for the relative errors of energy and
momentum vs. the time coordinate $x^0$ (we use the convention
$c=1$ throughout the paper).
\begin{figure}[thb]
\centerline{\includegraphics[scale=1.2
]{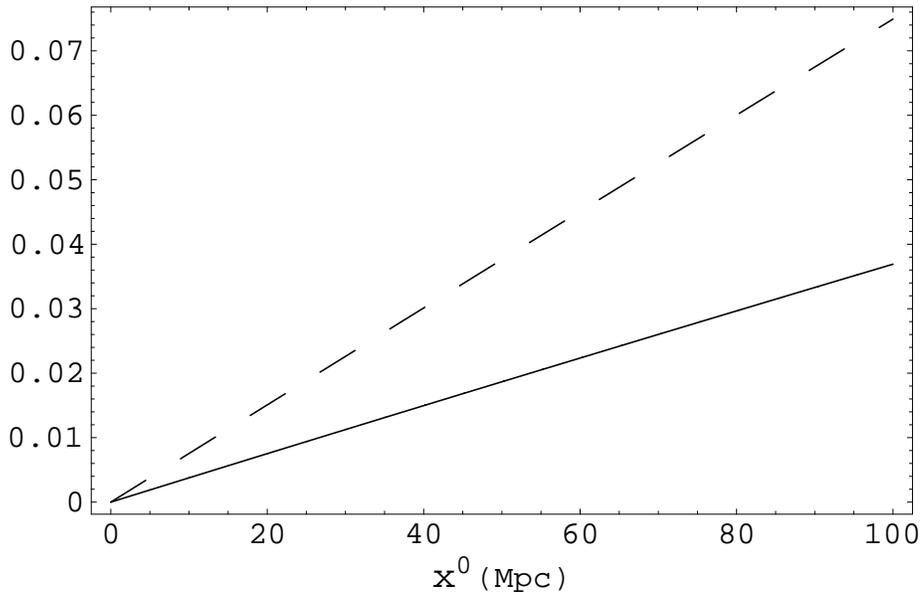}}
\caption{
The solid line describes the relative error of one-dimensional
momentum for a free particle moving in the perturbed dS universe,
while the dashed line describes its relative error of energy.}
\end{figure}

These results show clearly that, as a feasible approximation, we
can substitute energy and momentum in the empty dS universe for
the counterparts in perturbed one provided that the distance
between source of UHECRs and the earth is less than 100Mpc.

\section{Kinematics around dS spacetime}
In the asymptotically dS spacetime with homogeneous matter one can
still define the five dimensional angular momentum $M^{\mu\nu}$ of
a free particle with mass $m_0$ as the form
\begin{equation}\label{angular}
\hat{M}^{\mu\nu}=m_0\left(\hat{\xi}^{\mu}\frac{d\hat{\xi}^{\nu}}{ds}-\hat{\xi}^{\nu}
\frac {d\hat{\xi}^{\mu}}{ds}\right)~,
\end{equation}
where $s$ is a parameter along the geodesic. In the Universe,
there is no translation invariance and so that one can not
introduce a momentum vector. However, it should be noticed that,
at least somehow, one may define formally a counterpart of the
$4-$momentum $P$ for a free particle in the Universe as
\begin{equation}
\label{momentum} \hat{P}^{a}:= R^{-1}
\hat{M}^{4 a}=m_0\sigma^{-1}\frac{dx^{ a}}{ds}~. 
\end{equation}
In the same manner, the counterparts of the four dimensional
angular momentum $\hat{J}^{ab}$ can be assigned as $\hat{J}^{ab}:=
\hat{M}^{ab}$.

It is easy to show that $\xi^0(:=\sigma(x,x)^{-1/2}x^0)$ is
invariant under the spatial transformations. Thus, we can say  two
spacelike events are simultaneous if they satisfy
\begin{equation}
\label{simultaneous} \sigma(x,~x)^{-\frac{1}{2}}x^0=\xi^0={\rm
const.}~.
\end{equation}
Therefore, it is convenient to discuss physics of the
asymptotically dS spacetime in the coordinate $(\xi^0,~x^\alpha)$.
In this coordinate, the metric can be rewritten into the form
\begin{equation}
ds^2=\frac{d\xi^0d\xi^0}{1+\lambda\xi^0\xi^0}-(1+\lambda\xi^0\xi^0)(1+2\epsilon
f) \left[\frac{d\rho^2}{(1+\lambda\rho^2)^2}
+\frac{\rho^2}{1+\lambda\rho^2}d\Omega^2\right]~,
\end{equation}
where $\rho^2:=\sum\limits_{\alpha=1}^3{x^{\alpha}x^{\alpha}}$ and
$d\Omega^2$ denotes the metric on 2-dimensional sphere $S^2$.

The Klein-Gordon equation describes motions of a scalar field
$\phi(x)$,
\begin{equation}
\label{scalar} \left(\Box-m_0^2\right)\phi(x)=0~,
\end{equation}
where $\Box$ is the d'Alembertian operator defined as
\begin{equation}
\label{invariant}
\Box:=-\frac{1}{\sqrt{-g}}\partial_a(\sqrt{-g}g^{ab}\partial_b)~.
\end{equation}
The Dirac equation for the spinor field $\Psi(x)$ has the
form\cite{newadded}
\begin{equation}
\label{spinor}
\left[-i\gamma^{a}(\partial_{a}-\Gamma_{a})+m_0\right]\left(\begin{array}{c}\psi_1(x)\\
\psi_2(x)\end{array}\right)=0~,
\end{equation}
where $\Gamma_{a}$ is the Ricci rotational
coefficient\cite{wulixuebao-2}. For free spin $1/2$ particles, it
is easy to check that the components $\psi_{\alpha}(x)$ of a spinor
satisfy the relation
\begin{equation}
\label{spinor-1} \left(\Box-m_0^2\right)\psi_{ \alpha}(x)=0~,~~~~~ \alpha=1,~2.
\end{equation}
In Lorentz gauge, one can simplify the Maxwell equation 
without source as
\begin{equation}
\label{vector} \Box A^{a}=0~.
\end{equation}
From Eqs. (\ref{scalar}), (\ref{spinor-1}) and (\ref{vector}), one
knows that scalar fields and components of spinor and vector
fields can be described uniformly as ($m_0=0$ for vector field)
\begin{equation}\label{K-G-equation}
\left(\frac{1}{\sqrt{-g}}\partial_a(\sqrt{-g}g^{ab}\partial_b)-m_0^2
 \right)\Phi(\xi^0,x^{i})=0~.
\end{equation}

In the coordinates $(\xi^0,~x^{i})$, one can rewrite the
d'Alembertian operator as the following form
\begin{eqnarray}
\displaystyle\frac{1}{\sqrt{-g}}\partial_a(\sqrt{-g}g^{ab}\partial_b)
 &=&-\left(
1+\lambda\xi^0\xi^0
\right)\partial_{\xi^0}^2-\left(4\lambda\xi^0+3\epsilon(1+\lambda\xi^0\xi^0)\frac{df}{d\xi^0}\right)
\partial_{\xi^0}  \nonumber \\
\displaystyle&&+(1-2\epsilon f)\left( 1+\lambda\xi^0\xi^0
\right)^{-1}\left( 1+\lambda\rho^2
\right)^2\left[ \partial^2_\rho+2\rho^{-1}\partial_\rho \right] \\
\displaystyle&&+(1-2\epsilon f)\left( 1+\lambda\xi^0\xi^0
\right)^{-1}\left( 1+\lambda\rho^2 \right)\rho^{-2}
\partial^2_{\bf u} ~, \nonumber
\end{eqnarray}
where $\lambda=R^{-2}$, ${\bf u}{\bf u}^{\prime}=1$ and
$\partial^2_{\bf u}$ denotes the Laplacian operator on $S^2$.

To solve the equation of motion, one writes the field
$\Phi(\xi^0,x^{i})$ into the form
\[
\Phi(\xi^0,\rho,{\bf u})=T(\xi^0)U(\rho)Y_{lm}(\bf u)~.
\]
Thus, one transforms the equation of motion into
\cite{ds-paper-3,wulixuebao-2},
\begin{eqnarray}
&& \left[(1+\lambda\xi^0\xi^0)^2(1+2\epsilon
f)\partial^2_{\xi^0}+\left(4\lambda\xi^0(1+\lambda\xi^0\xi^0)(1+2\epsilon
f)+3\epsilon\frac{df}{d\xi^0}(1+\lambda\xi^0\xi^0)^2\right)
\partial_{\xi^0}\right.\nonumber\\ [0.3cm]
&&\left.~~~~~~~~~~~~~~~~~~~~~~~~~~~+m_0^2(1+2\epsilon
f)(1+\lambda\xi^0\xi^0)+(\varepsilon^2-m_0^2)\right]T(\xi^0)=0,
\nonumber \\[0.5cm]
&& \left[\partial^2_{\rho}+
\displaystyle\frac{2}{\rho}\partial_\rho-\left[\frac{m_0^2-\varepsilon^2}
{(1+\lambda\rho^2)^2}+\frac{l(l+1)}{\rho^2(1+\lambda\rho^2)}\right]
\right]U(\rho)=0, \nonumber\\[0.5cm]
&& \left[\partial^2_{\bf u}+l(l+1)\right]Y_{lm}({\bf u})=0,
\end{eqnarray}
where $Y_{lm}({\bf u})$ is the spherical harmonic function and
$\varepsilon$ is the constant from separating variables.

For the irrelevance of the expression of $T(\xi^0)$ to our
discussion, we can focus our attention on the last two equations.
The solutions for the radial equation of the field is
\begin{equation}
U(\rho)\sim\rho^l(1+\lambda\rho^2)^{k/2}F\left(\frac{1}{2}(l+k+1),~\frac{1}{2}
(l+k),~l+\frac{3}{2};~-\lambda\rho^2\right)~,
\end{equation}
where $k$ denotes the radial quantum number
$$k^2-2k-\lambda^{-1}(\varepsilon^2-m_0^2)=0~.$$
To be normalizable, the hypergeometric function in the radial part
of the wavefunction has to break off, leading to the quantum
condition
\omits{\begin{equation}
\fbox{$\frac{l+k}{2}=-n~,~~~~~(n\in \mathbf{N})~.$}
\end{equation}}
\be l+k = -2n , \qquad  (2n\in \mathbf{N}). \ee Then, one obtains
the dispersion relation for a free particle moving in the Universe
\begin{equation}
\label{dispersion2}
E^2=m_0^2+\varepsilon'^2+\lambda(2n+l)(2n+l+2)~,
\end{equation}
where the term $\varepsilon'^2$ denotes what is independent of the
parameters $n$ and $l$.

\section{UHECR threshold}
In this section, we investigate the UHECR threshold in the
covariant framework of kinematics in an asymptotically dS
spacetime set up in the preceding sections.

One considers the head-on collision between a soft photon of
energy $E_\gamma$, momentum ${\bf q}$ and a high energy particle
$m_1$ of energy $E_1$, momentum ${\bf p}_1$, which leads to the
production of two particles $m_2,~m_3$ with energies $E_2$, $E_3$
and momentums ${\bf p}_2$, ${\bf p}_3$, respectively. From the
energy and momentum conservation laws
\begin{equation}\label{threshold11}
\begin{array}{c}
E_1+E_\gamma=E_2+E_3~, \\[0.4cm]
p_1-q=p_2+p_3~.
\end{array}
\end{equation}
In the C. M. frame, $m_2$ and $m_3$ are at rest when the threshold
is reached, so they have the same velocity in the lab frame and
there exists the following relation
\begin{equation}
\label{p3p2} \frac{p_2}{p_3}=\frac{m_2}{m_3}~.
\end{equation}
It is convenient to use the approximate formulae of dispersion
relations (\ref{dispersion2}) for the soft photons and the ultra
high energy particles
\begin{eqnarray}\label{dispersion-app}
&&E_\gamma^2=q^2+\lambda_{\gamma}^{*}~,\\[0.4cm]
&&E_i=\sqrt{m^2_i+p^2_i+\lambda_i^{*}}\simeq
p_i+\frac{m^2_i}{2p_i}+\frac{\lambda^{*}_i}{2p_i}~,~~(i=1,~2,~3)~.
\end{eqnarray}
where
$~\lambda_{\gamma}^{*}:=\lambda(l_\gamma+2n_\gamma)(l_\gamma+2n_\gamma+2)
\approx\lambda {l_{\gamma}}(l_{\gamma}+2)~$ and
$~\lambda_{i}^{*}:=\lambda(l_i+2n_i)(l_i+2n_i+2)\approx\lambda
{l_{i}}(l_{i}+2)~$ with the conjecture that $~l \gg n~$.

The obtained threshold can be expressed as the form
\begin{equation}
\label{threshold0}\displaystyle E_{{\rm
~th},~\lambda}\simeq\frac{(m_{2}+m_{3})^2-m_{1}^2+\lambda_{2}^{*}
\left(1+\frac{m_3}{m_2}\right)+\lambda_{3}^{*}\left(1+\frac{m_2}{m_3}\right)-
\lambda_{1}^{*}}{2\left(E_\gamma+\sqrt{E_\gamma^2-\lambda_{\gamma}^{*}}\right)}~.
\end{equation}
The usual GZK threshold could be recovered when the parameter
$\lambda^*$, which is dependent on the cosmological constant, runs
to zero.

The conservation law of the angular momentum imposes a constraint
on the parameters $\lambda^*$,
$$
\lambda^{*}_{1} + \lambda^{*}_{\gamma} + 2\lambda{\bf
L}_1\cdot{\bf L}_{\gamma} = \lambda^{*}_{2} + \lambda^{*}_{3} +
2\lambda{\bf L}_2\cdot{\bf L}_3~.
$$
Making use of the relation, one can rewrite the $\lambda^*$
dependent terms of the threshold as the following
\begin{equation}
\frac{\lambda^{*}_{2}\frac{m_3}{m_2}+\lambda^{*}_{3}\frac{m_2}{m_3}+\lambda^{*}_{\gamma}
+2\lambda{\bf L}_1\cdot{\bf L}_{\gamma}-2\lambda{\bf L}_2\cdot{\bf
L}_3}
{2\left(E_\gamma+\sqrt{E_\gamma^2-\lambda_{\gamma}^{*}}\right)}~.
\end{equation}
If $\lambda^{*}_2$ and $\lambda^{*}_3$  take value of the same
order with $\lambda^{*}_{\gamma}$ (less than the square of energy
of a soft photon), the $\lambda^*$ dependent terms can be
omitted\cite{0307439}.  We will investigate the case of
$\lambda^{*}_2+\lambda^{*}_3 \gg \lambda^{*}_{\gamma}~$, and the
threshold $(\ref{threshold0})$ is of the form
\begin{equation}
\label{threshold}\displaystyle E_{{\rm
~th},~\lambda}\simeq\frac{(m_{2}+m_{3})^2-m_{1}^2+
\lambda^{*}_{2}\frac{m_3}{m_2}+\lambda^{*}_{3}\frac{m_2}{m_3}-
2\lambda{\bf L}_2\cdot{\bf L}_3}{2\left(E_\gamma+
\sqrt{E_\gamma^2-\lambda_{\gamma}^{*}}\right)}~.
\end{equation}

Now, we can study the photopion production processes of the UHECR
interaction with the CMBR
$$p+\gamma\rightarrow p+\pi~.$$
The corresponding threshold for this process is given by
\begin{equation}
\label{UHECR-threshold}\displaystyle E^{\rm UHECR}_{{\rm
~th},~\lambda}\simeq\frac{(m_{N}+m_{\pi})^2-m_{N}^2+
\lambda^{*}_{N}\frac{m_{\pi}}{m_N}+\lambda^{*}_{\pi}\frac{m_N}{m_{\pi}}-
2\lambda{\bf L}_{N}\cdot{\bf L}_{\pi}}{2\left(E_\gamma+
\sqrt{E_\gamma^2-\lambda_{\gamma}^{*}}\right)}~.
\end{equation}
To show the behavior of the threshold in the $\lambda^*$-parameter
space clearly, we should discuss some limit cases in detail.

In the case that the out-going nucleon has zero angular momentum,
the threshold $(\ref{UHECR-threshold})$  reduces  as
\begin{equation}
\label{UHECR-pi}\displaystyle E^{\rm UHECR}_{{\rm
~th},~\lambda,\pi}\simeq\frac{(m_{N}+m_{\pi})^2-m_{N}^2+
\lambda^{*}_{\pi}\frac{m_N}{m_{\pi}}}{2\left(E_\gamma+
\sqrt{E_\gamma^2-\lambda_{\gamma}^{*}}\right)}~.
\end{equation}
we provide a plot for the dependence of the threshold $E^{\rm
UHECR}_{{\rm th},~\lambda,\pi}$ on the cosmological constant and
angular momentums (the in-coming photon and out-going pion) in
Fig. 2.
\begin{figure}[hpb]
\centerline{\includegraphics[scale=1.2
]{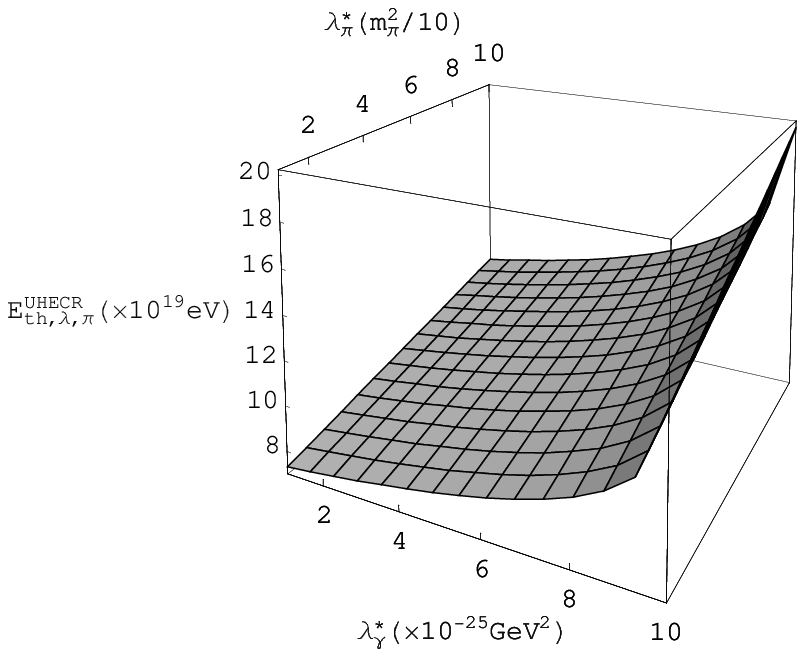}}
\caption{
The cosmological constant and angular momentums (of in-coming soft
photon and out-going pion) dependence of the threshold $E^{\rm
UHECR}_{{\rm th},~\lambda,\pi}$ in the interaction between the
UHECR protons and the CMBR photons ($\lambda^{*}_{\pi}$ in units
of $m_{\pi}^2/10$)~.}
\end{figure}

In the case that the out-going pion has zero angular momentum, the
UHECR threshold takes the form
\begin{equation}
\label{UHECR-pn}\displaystyle E^{\rm UHECR}_{{\rm
~th},~\lambda,N}\simeq\frac{(m_{N}+m_{\pi})^2-m_{N}^2+
\lambda^{*}_{N}\frac{m_{\pi}}{m_{N}}}{2\left(E_\gamma+
\sqrt{E_\gamma^2-\lambda_{\gamma}^{*}}\right)}~.
\end{equation}
We present a plot for the dependence of the threshold $E^{\rm
UHECR}_{{\rm th},~\lambda,N}$ on the cosmological constant and
angular momentums (the in-coming soft photon and out-going
nucleon) in Fig. 3.
\begin{figure}
\centerline{\includegraphics[scale=1.2
]{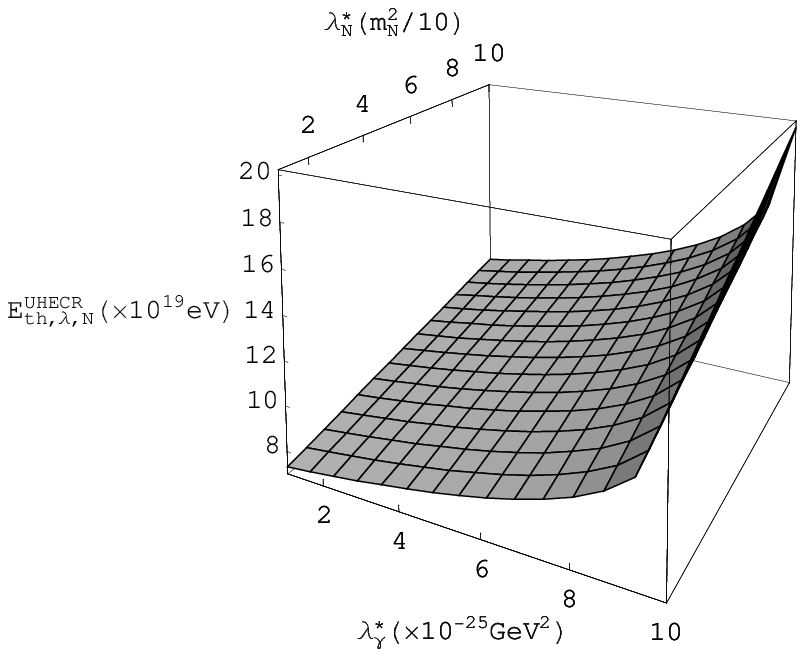}}
\caption{
The cosmological constant and angular momentums (of the in-coming
soft photon and out-going nucleon) dependence of the threshold
$E^{\rm UHECR}_{{\rm th},~\lambda,N}$ in the interaction between
the UHECR protons and the CMBR photons ($\lambda^{*}_{N}$ in units
of $m_{N}^2/10$)~.}
\end{figure}

Finally, if the out-going pion and nucleon have the same angular
momentum, the UHECR threshold can be expressed as the following
form
\begin{equation}
\label{UHECR-pnpi}\displaystyle E^{\rm UHECR}_{{\rm
~th},\lambda,N\pi}\simeq\frac{(m_{N}+m_{\pi})^2-m_{N}^2+
\lambda^{*}_{N}(\frac{m_{N}}{m_{\pi}}+\frac{m_{\pi}}{m_{N}}-2)}{2\left(E_\gamma+
\sqrt{E_\gamma^2-\lambda_{\gamma}^{*}}\right)}~.
\end{equation}
We provide a plot for the dependence of the threshold $E^{\rm
UHECR}_{{\rm th},~\lambda,N\pi}$ on the cosmological constant and
angular momentums (the in-coming soft photon and out-going nucleon
and pion) in Fig. 4.

From the above discussion, one can conclude that a tiny but
nonzero cosmological constant may provide indeed sufficient
corrections to the primary predicted threshold\cite{GZK}. For the
observed cosmological constant (which is around the level of
$10^{-85}{\rm GeV}^2$), if the CMBR possesses a quantum number
$l_{\gamma}$ of the order of $10^{30}$, the threshold will be
above the energies of all those observed UHECR particles.  The
predicted threshold should be upgraded to a more reasonable level.
Now we can say that a possible origin of the cosmic ray threshold
anomaly has been achieved. It is the cosmological constant that
increases the GZK cut-off to a level above the observed UHECR
events.
\begin{figure}
\centerline{\includegraphics[scale=1.2
]{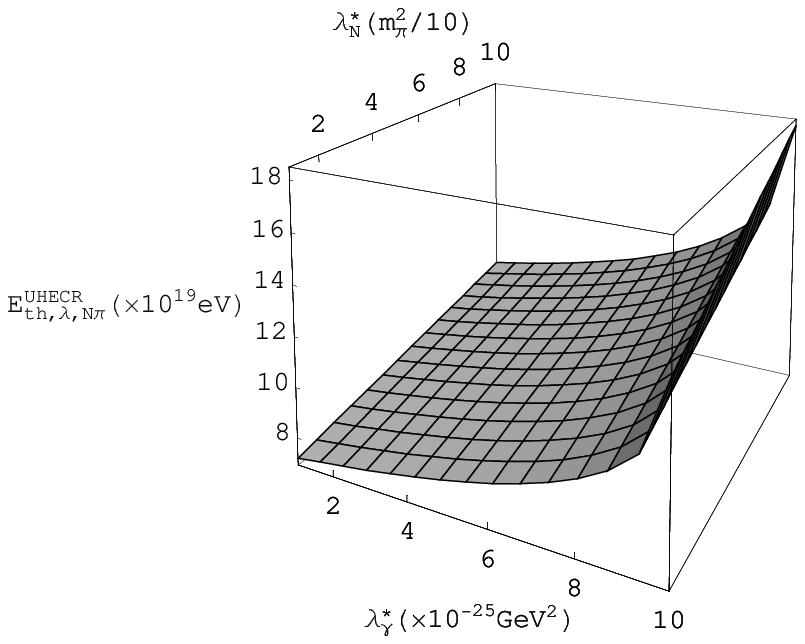}}
\caption{
The cosmological constant and angular momentums (of the out-going
nucleon and pion) dependence of the threshold $E^{\rm UHECR}_{{\rm
th},~\lambda,N\pi}$ in the interaction between the UHECR protons
and the CMBR photons ($\lambda^{*}_{N}=\lambda^{*}_{\pi}$ in units
of $m_{\pi}^2/10$)~.}
\end{figure}

\section{Conclusions and remarks}
In this paper, we have showed a cosmological scenario of constant
curvature spacetime with a homogeneous density perturbation. This
is in agreement with the astronomical observations on supernovae
and CMBR  that about two thirds of the whole energy in the
Universe is from dark energy and the lowest order description of
the Universe may be a de Sitter spacetime.  The matter world was
dealt with as a perturbation around the de Sitter background
within the framework of standard cosmology.

Kinematics in an asymptotically dS spacetime and, in particular,
the perturbation around the dS background was presented.  The
exact conservation law of momentum and angular momentum in dS is
violated by virtue of the additional homogeneous density
perturbation. Numerical simulations show that, within the range
that we are interested, the violation of the conservation law is
small. A general form of dispersion relation for free particles
moving in the Universe was obtained.

The perturbation around dS spacetime has been used to discuss  the
UHECR threshold anomaly.  We obtained explicitly the corrections
of the GZK threshold for the UHECR particles interacting with soft
photons, which are dependent on the cosmological constant. We
showed how the threshold varies with a positive cosmological
constant and additional degrees of freedom of the angular
momentums of interacting particles. It should be noticed that, for
a positive cosmological constant, the theoretic threshold tends to
be above the energies of all the observed events. Thus, we may
conclude that the tiny but nonzero cosmological constant is a
possible origin of the threshold anomaly of the UHECR.

\vspace{0.5cm}

\noindent{\large\bf Acknowledgement:}\\
We are grateful to Prof. H. Y. Guo  C. J. Zhu, and Z. Xu for
useful discussion. This work is partly supported by the Natural
Science Foundation of China under Grants No. 10375087, 10375072,
90403023. One of us (C. B. Guan) is supported by grants through
the ICTS (USTC) from the Chinese Academy of Sciences.

\end{document}